\documentclass{elsart}
\usepackage{times}
\usepackage{courier}
\usepackage{algorithm}
\usepackage{algorithmic}
\usepackage{natbib}
\usepackage{epsfig}
\usepackage{amssymb}

\def\dj{\hbox{d\kern-0,347em \vrule width0,3em height1,252ex
depth-1,21ex \kern0,051em}}
\begin{document}

\begin{frontmatter}

\title{Scaling of Fracture Strength in Disordered Quasi-Brittle Materials}

\author[1]{Phani Kumar V.V. Nukala}
\author[1]{Sr{\dj}an \v{S}imunovi\'{c}}
\address[1]{Computer Science and Mathematics Division, 
Oak Ridge National Laboratory, Oak Ridge, TN 37831-6164, USA}

\begin{abstract}
This paper presents two main results. The first result indicates that
in materials with broadly distributed microscopic heterogeneities, 
the fracture strength distribution corresponding to the peak load of the material response  
does not follow the commonly used Weibull and (modified) Gumbel distributions.
Instead, a {\it lognormal} distribution describes more adequately the fracture strengths 
corresponding to the peak load of the response. 
Lognormal distribution arises naturally as a consequence of multiplicative nature of large number of 
random distributions representing the stress scale factors necessary to break 
the subsequent "primary" bond (by definition, an increase in applied stress is required to break a
"primary" bond) leading up to the peak load.
Numerical simulations based on two-dimensional triangular and diamond
lattice topologies with increasing system sizes
substantiate that a {\it lognormal} distribution represents an excellent fit 
for the fracture strength distribution at the peak load.
The second significant result of the present study is that, 
in materials with broadly distributed microscopic heterogeneities,
the mean fracture strength of the 
lattice system behaves as $\mu_f = \frac{\mu_f^\star}{(Log L)^\psi} ~+~ \frac{c}{L}$, 
and scales as $\mu_f \approx \frac{1}{(Log L)^\psi}$ 
as the lattice system size, $L$, approaches infinity.
\end{abstract}


\end{frontmatter}

\section{Introduction}
It is well known that fracture properties and breakdown behavior 
are very sensitive to the 
microstructural details of the material \cite{lawn}. 
In ductile materials, grain boundaries and 
second-phase particles form the important microstructural details for fracture 
and damage evolution. In quasi-brittle materials such as ceramics, the fracture properties
are usually dominated by the size and spatial distribution of microcracks, which
are often the artifacts of material processing techniques. 
The fracture strength distribution of a quasi-brittle material is significantly 
influenced by the distribution of microcracks. Probabilistic life design 
methodologies, "useful" service life predictive models, and failure risk analysis 
of a structural material component utilize fracture strength distributions in 
assessing the safety and reliability of the material component.

\par
Traditionally, Weibull 
and (modified) Gumbel distributions based on "weakest-link" 
approach have been widely used to 
describe the strength of brittle materials. These distributions naturally arise from 
the extreme-value statistics if one assumes the following conditions for the 
distribution of defect cluster sizes in a randomly diluted network \cite{peterlik1}: 
(1) defect clusters are independent of each other, i.e., they do not interact with one another;
(2) system failure is governed by the "weakest-link" hypothesis, and 
(3) there exists a critical defect cluster size below which the system does not fail, and 
it is possible to relate the critical size of a defect cluster to the material strength. 
Moreover, if the defect cluster size distribution is described by a power-law, then 
the fracture strength obeys Weibull distribution, whereas an exponential 
defect cluster size distribution leads to the Gumbel distribution for fracture strengths.
However, in heterogeneous materials with broad distribution of disorder, Weibull 
and Gumbel distributions may not adequately represent the fracture strengths 
corresponding to the peak load response. 
There are two main reasons behind this inadequacy. 

\par
First, in the "weakest-link" hypothesis, 
the fracture strength of a randomly diluted lattice 
system is determined by the presence of few critical defect 
clusters, 
and is defined as the stress required for breaking the very first
"weakest-link" in the system. In materials with broad disorder, the breaking of the 
very first bond ("weakest-link") does not usually lead to the entire system failure,
and hence the fracture strength distribution based on the very first bond failure 
may not be applicable.
In addition, the evolution of initial distribution of defect clusters, i.e., 
subsequent bond failure is controlled not only by "weakest" bonds with smallest 
thresholds but also by the stress concentration and shielding effects around the 
defect clusters.
Consequently, at the peak load,
the defect cluster size distribution that evolved under the applied stress field
may be quite different from the initial defect cluster size distribution.
Thus, unless the defect cluster size distributions
at the peak load and at the very first bond failure have the same form,
it is
highly unlikely that the fracture strength distributions corresponding to the peak load
and the first bond failure would follow the same distribution.

\par
Secondly, if the defect cluster size distribution at the peak load exhibits 
multi-modal behavior then the corresponding fracture strength distribution 
is not adequately described by unimodal distributions such as Weibull and Gumbel distributions.
In the presence of 
experimentally observed multiple
defect populations, the tail cannot be fitted by a power law with a single power law 
exponent
due to the superposition of different defect cluster size distributions. 
These multi-modal defect cluster size populations result in 
multi-modal strength distributions \cite{poloniecki,jayatilaka,sigl,lissart,peterlik1}. 
Experimentally, bimodal Weibull
strength distributions were observed in various brittle materials including
carbon \cite{helmer} and silicon carbide fibers
\cite{lissart}, and for certain ceramics \cite{orlovskaja}. The effect of
multiple flaw (defect) populations on fracture strength distribution
is also examined experimentally in certain ceramic materials
\cite{chao,orlovskaja1,peterlik,zhang}.  
In general at the peak load, the broken bond cluster distribution follows
generalized Gamma-distribution and in this case, 
neither the Weibull nor the Gumbel distributions
fit the fracture strength distribution accurately \cite{sahimi933,lissart,peterlik1}.

\par
This paper presents two main results. The first result indicates that, in the case of 
materials with broadly distributed heterogeneities, a lognormal
distribution represents the fracture strength of the macroscopic system more adequately 
than the conventional (modified) Gumbel and Weibull distributions. The second result 
indicates that the mean fracture strength decreases with increasing lattice system size, $L$,  
and scales as $\frac{1}{(Log L)^\psi}$ in the 
thermodynamic limit.

\par
The paper is organized as follows. Section 2, presents the theoretical derivation 
for the fracture strength distribution of a lattice system. 
In Section 3, numerical simulations 
using two-dimensional triangular and diamond (square lattice inclined at 
45 degrees between the bus bars) lattice networks are presented. Section 4
presents the validation results and a comparison study between the lognormal 
distribution for fracture strengths derived in this study versus the traditionally 
used Weibull and Gumbel distributions.

\section{Fracture Strength of a Discrete Lattice System}
Progressive damage evolution leading to failure of disordered
quasi-brittle materials has been
studied extensively using various types of discrete lattice models
\cite{hansen001,herrmann90,sahimi98,chakrabarti}.
Electrical fuse/breakdown models \cite{duxbury86,duxbury87,duxbury88,deArcangelis88,kahng,roux91}, central-force models \cite{deArcangelis85,sahimi86,beale88,feng,hansen89B,hansen89C,sahimi931}, bond-bending
models \cite{kantor,sahimi932}, and beam-type models \cite{herrmann90,herrmann89}
have been used in combination with disorder either in the
elastic constants, threshold values or in the random dilution of the bonds to
model damage evolution in brittle materials.
The reader is referred to \cite{chakrabarti,hansen001,herrmann90,sahimi98}
and the references therein for a comprehensive review of modeling quasi-static progressive
damage evolution in brittle materials using discrete lattice networks.
Although discrete lattice
models do not describe the specific behavior of any real material, they
incorporate the essential ingredients of a breakdown process, namely,
the initial material disorder and the redistribution of forces due to damage
evolution. In this respect, any realistic progressive damage evolution model that
describes the behavior of real materials should be capable of reproducing the
behavior of these idealized discrete lattice models \cite{delaplace}.

\par
The essential features of discrete lattice models are disorder, elastic response 
characteristics, and a breaking rule for each of the bonds in the lattice.
The elastic and breaking response characteristics of each bond in the lattice
correspond to the mesoscopic response of the material.
The elastic response of the individual bonds is typically described by
electrical fuse models, central-force (spring) models, bond-bending spring models,
and beam-type models. The quenched disorder in the system is introduced either in the
elastic constants, in the threshold values, or by the random dilution of bonds.
The breaking of a bond occurs irreversibly
when the applied action (stress or displacement) across the
bond exceeds the breaking threshold. Various types of breaking rules have been adopted
in the literature depending on the elastic response characteristics of individual bonds.
When an external action (displacement or force) on the lattice is slowly increased,
the individual bonds in the lattice will break irreversibly one after another until the
system falls apart. 
It is supposed that
successive fuse failures leading ultimately to the failure of lattice
 system is similar to the breakdown of quasi-brittle materials.

\par
In this study, we consider a discrete lattice system in which all the bonds are
intact at the beginning of the analysis and the damage is accumulated progressively
by breaking one bond at a time until the entire lattice system falls apart.
This approach is in contrast with the
earlier works, wherein either randomly diluted lattice systems close to the
percolation threshold were considered or the stress required to break the very first
bond was defined as the fracture strength of the lattice system. By using either of these
methodologies, it is possible to analysize very large lattice systems of size $1000 \times 1000$.
However, the defect cluster distribution obtained by randomly diluting the bonds of a lattice systems is
quite different from that obtained by sequentially breaking one bond at a time.
This is because,
the evolution of initial distribution of defect clusters 
is controlled not only by "weakest" bonds with smallest
thresholds but also by the stress concentration effects around the
defect clusters. In this paper, 
we define the fracture strength of a lattice system as the stress corresponding 
to the peak load of the lattice system response.

\par
\begin{figure}[hbtp]
\centerline{\includegraphics[width=12cm]{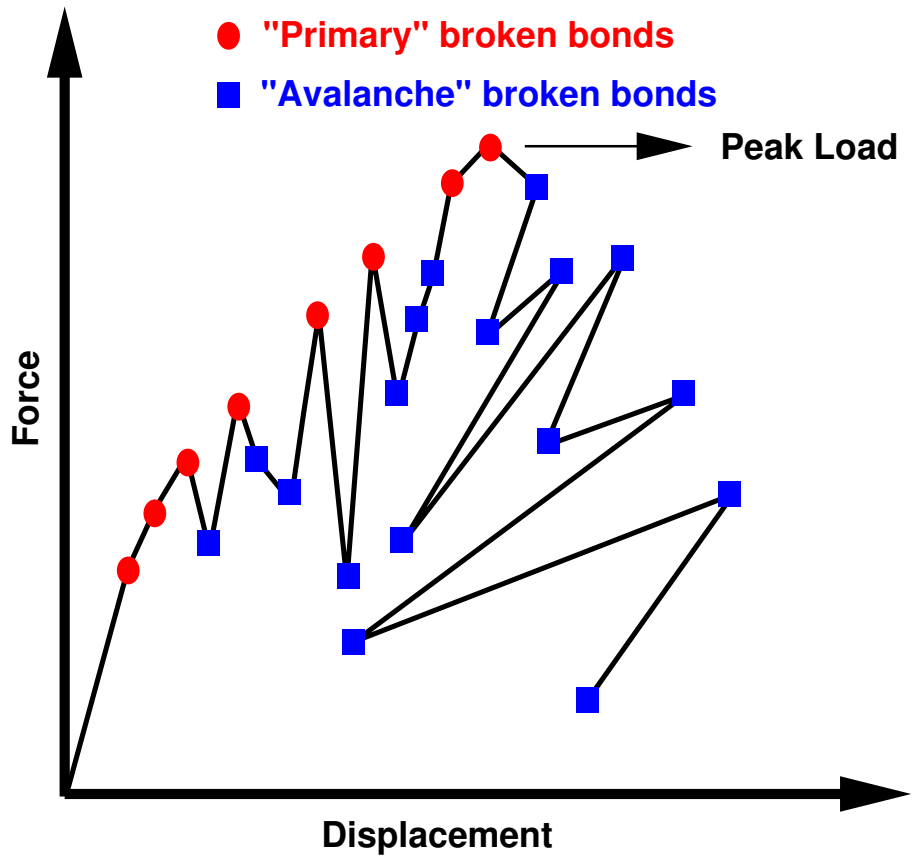}}
\caption{A typical lattice system response indicating "primary" and "avalanche"
broken bonds.}
\label{fig1}
\end{figure}

Consider a lattice system with a total number of bonds, $N_{el}$, subjected to a 
stress controlled loading. Let 
${\mathcal N} = \{1,2,3,\cdots,N_{el}\}$ denote the set of individual bonds in the 
lattice system. After breaking $k$ number of bonds, let ${\mathcal S}_{k}^{b}$
denote the set of all the $k$ number of bonds that were 
broken. 
A broken bond is considered a "primary" bond if an increase in the 
applied stress is necessary to break that bond. 
Similarly, let ${\mathcal S}_{k}^{a}$ denote the set of bonds that will be broken 
("avalanche" bonds) without any further increase in the externally applied stress, 
given the set of broken bonds 
${\mathcal S}_{k}^{b}$ at an applied stress $\sigma$. 
Figure \ref{fig1} illustrates this concept using a 
typical lattice system response. Once a "primary" bond is broken, 
subsequent breaking of the "avalanche" bonds continues until we encounter 
another "primary" bond, i.e., until it is required to increase the applied stress level
to break a bond. If we do not encounter a "primary" bond, it means that we reached the 
peak load of the lattice system.
The dimension of the 
set ${\mathcal S}_{k}^{a}$ represents the avalanche size, and is given by $a_k$.
Let ${\mathcal S}_{k}^{u}$ denote the set of unbroken bonds remaining in the lattice system 
after $k$ number of bonds and the elements of the avalanche set ${\mathcal S}_{k}^{a}$ 
have already been broken. It should be noted that the set 
${\mathcal S}_{k}^{u}$ also includes the set of bonds ${\mathcal D}_k^u$ 
called the {\it dangling} or {\it dead ends}
that do not carry any stress even though they are not broken. The sets 
${\mathcal S}_{k}^{b}$, ${\mathcal S}_{k}^{a}$ and ${\mathcal S}_{k}^{u}$ are mutually disjoint
and collectively exhaustive such that
\begin{equation}
\left.\begin{array}{ccccc}
{\mathcal N} & = & {\mathcal S}_{k}^{b} \oplus {\mathcal S}_{k}^{a} \oplus {\mathcal S}_{k}^{u} \\
{\mathcal S}_{k}^{b} \cap {\mathcal S}_{k}^{a} & = & \varnothing \\
{\mathcal S}_{k}^{a} \cap {\mathcal S}_{k}^{u} & = & \varnothing \\
{\mathcal S}_{k}^{b} \cap {\mathcal S}_{k}^{u} & = & \varnothing \\
\end{array} \right\}
\end{equation}
where $\varnothing$ denotes the null set, and $\oplus$ denotes the additive sum of disjoint sets.
Assume that for each non-empty set ${\mathcal S}_{k}^{b}$, a set of non-negative 
stress concentration
constants ${\mathcal L}_k = \{\lambda_{i}({\mathcal S}_{k}^{b} \oplus 
{\mathcal S}_{k}^{a}), i \in {\mathcal S}_{k}^{u}\}$
are defined. These constants 
represent the stress concentration coefficients for all the elements of
the set of unbroken bonds ${\mathcal S}_{k}^{u}$ when all the elements of the 
sets ${\mathcal S}_{k}^{b}$ and ${\mathcal S}_{k}^{a}$ are removed from the lattice system. 
Denote the stress concentration coefficient $\lambda_{i}({\mathcal S}_{k}^{b} \oplus 
{\mathcal S}_{k}^{a})$ to be equal to zero for all the bonds $\{i \in {\mathcal D}_{k}^{u}\}$.
These stress concentration coefficients ${\mathcal L}_k$ implicitly 
define the redistribution of stress within the lattice system after all the 
elements in the set ${\mathcal S}_{k}^{b} \oplus {\mathcal S}_{k}^{a}$ have been removed.

\par
Let ${\mathcal M}_k$ denote the set of non-negative constants such that 
${\mathcal M}_k = \{\mu_i = 1/\lambda_i: i \in ({\mathcal S}_{k}^{u} - {\mathcal D}_{k}^{u})\}$.
Also, let ${\mathcal X}_{k}^{th} = \{x_i = \mu_i \sigma_i^{th}: 
\mu_i \in {\mathcal M}_k, \sigma_i^{th} \sim P_0^{th}(\sigma) ~\mbox{and}~ 
i \in ({\mathcal S}_{k}^{u} - {\mathcal D}_{k}^{u})\}$ denote the 
set of modified breaking thresholds (similar to annealed disorder) 
for all the bonds in the set $({\mathcal S}_{k}^{u} - {\mathcal D}_{k}^{u})$
after the bonds in the set ${\mathcal S}_{k}^{b} \oplus {\mathcal S}_{k}^{a}$ have been broken.
In the above description, $P_0^{th}(\sigma)$ denotes the cumulative probability 
distribution of the breaking thresholds at $0^{th}$ state, i.e., before any of the
bonds in the lattice system are broken. However, as the bonds are broken 
successively, the distribution of breaking thresholds $P_k^{th}(\sigma)$ of the set 
of elements ${\mathcal X}_{k}^{th}$ changes gradually from its initial distribution
$P_0^{th}(\sigma)$ of the elements in ${\mathcal N}$. 

\par
The conditional probability that a bond from the set ${\mathcal S}_{k}^{u}$ breaks given 
that the bonds in the set ${\mathcal S}_{k}^{b} \oplus {\mathcal S}_{k}^{a}$ have been 
broken, i.e., for breaking 
the $(k + a_k + 1)^{th}$ bond in the lattice system given that $(k + a_k)$ have been 
broken already, is given by
\begin{equation}
f_{(k + a_k + 1)} = P_k^{th}(\sigma_{(k + a_k + 1)}^{min})
\end{equation}
where $f_{(k + a_k + 1)}$ denotes the conditional probability of breaking the 
$(k + a_k + 1)^{th}$ bond, and $\sigma_{(k + a_k + 1)}^{min} = inf({\mathcal X}_{k}^{th})$. 
Note that $\sigma_{(k + a_k + 1)}^{min}$ denotes the minimum externally applied stress 
that is necessary to break the $(k + a_k + 1)^{th}$ bond. 
The conditional probabilities for 
breaking any of the bonds within the set ${\mathcal S}_{k}^{a}$, given that 
all the bonds in the set ${\mathcal S}_{k}^{b}$ are broken is equal to one, i.e., all the 
bonds in the set ${\mathcal S}_{k}^{a}$ break without any further increase in stress 
after breaking the $k^{th}$ bond. Hence,
\begin{equation}
f_{(k+j)} = 1  ~~\forall ~j = \{1,2,\cdots,a_k\}
\end{equation}
Thus, the probability of breaking the 
$(k + a_k + 1)$ bond in the lattice system is
\begin{eqnarray}
F_{(k + a_k + 1)} & = & f_{(k + a_k + 1)} ~F_{(k + a_k)} \nonumber \\
& = & f_{(k + a_k + 1)} ~f_{(k + a_k)} ~f_{(k + a_k - 1)} \cdots f_{(k + 1)} ~F_{(k)} 
\nonumber \\
& = & f_{(k + a_k + 1)} ~F_{(k)} \label{Fkak1}
\end{eqnarray}
Let ${\mathcal B}$ denote an ordered set of indices, where each index refers to the sequential 
number of the broken bond for which an increase in the applied stress is required to 
break the bond. Then, this set contains the indices such as $k$ and $(k + a_k + 1)$
as its elements, and is written in a general way as
\begin{equation}
{\mathcal B} = \{b_{j+1} = (b_{j} + a_j + 1): ~b_0 = 0; ~a_0 = 0; ~j \in \{0,1,2,\cdots\}\} \label{eqb}
\end{equation}
In Eq. (\ref{eqb}), $a_j$ refers to the avalanche size after breaking the $b_j$ indexed 
bond. It should be noted that the set ${\mathcal B}$ maps the sequential number of the 
"primary" broken bond to the sequential number of the broken bond. That is, 
$b_j = {\mathcal B}(j)$, where $j = \{1,2,3,\cdots\}$ refers to the sequential number of the 
"primary" broken bond and $b_j$ refers to the sequential number of the broken bond.
With this notation in hand, Eq. (\ref{Fkak1}) can be written 
recursively as
\begin{eqnarray}
F_{(k + a_k + 1)} & = & \prod_{j \in {\mathcal B}} f_{(j)} \label{Fkprod}
\end{eqnarray} 
Similarly, let ${\mathcal A}$ denote an ordered set of externally applied stress values 
that are required to break the "primary" bonds, i.e.,
\begin{equation}
{\mathcal A} = \{\sigma_j^{min}: ~j \in {\mathcal B}\} \label{eqa}
\end{equation}
Now, consider the set of scale factors, ${\mathcal G}$, defined as
\begin{equation}
{\mathcal G} = \{g_{b_j} = \frac{\sigma_{b_j}^{min}}{\sigma_{b_{j-1}}^{min}}: ~b_j = {\mathcal B}(j); ~b_{j-1} = {\mathcal B}(j-1); g_1 = 1; ~j = \{2,3,\cdots\}\} \label{eqg}
\end{equation}
Since the elements of the set ${\mathcal A}$ depend not only on the initial distribution of 
breaking thresholds but also on the stress concentration factors around the broken 
bond clusters, these stress levels, $\sigma_j^{min}$, at which the "primary" bonds break
can be considered as independently distributed random variables. Consequently, 
the elements of the set ${\mathcal G}$ are also independently distributed random variables.
Using Eqs. (\ref{eqa}) and (\ref{eqg}), the stress required to break the 
$(k + a_k + 1)^{th}$ bond, $\sigma_{(k + a_k + 1)}^{min}$, can be expressed as a 
product of independently distributed random scale factors, i.e.,
\begin{equation}
\sigma_{(k + a_k + 1)}^{min} = \left(\prod_{j \in {\mathcal B}} g_j\right) ~\sigma_{1}^{min} \label{sigprod}
\end{equation}
where $\sigma_{1}^{min}$ is the stress required to break the first bond.

\par 
In lattice systems with broadly distributed breaking thresholds, 
the cumulative avalanche sizes in the regime up to the peak load of the 
material response are negligible compared with the total number of bonds broken 
up to the peak load. Consequently, 
the dimension (cardinality) of the set ${\mathcal G}$ with independently distributed 
scale factors $g_j$, $j \in {\mathcal B}$, is approximately $O(n_p)$, where $n_p \approx O(L^{1.8})$ 
is the number of broken bonds up to the peak load. 
By virtue of the central limit theorem, which states that 
the product of a large number of independent 
factors, none of which dominates the product, will tend to the lognormal distribution
regardless of the distributions of the individual factors involved in the product, 
we have, $\mbox{Prob}[\sigma_{(k + a_k + 1)}^{min} \le \sigma] \sim LN$ 
as $\mbox{dim}({\mathcal B}) \rightarrow \infty$, 
where $LN$ denotes lognormal distribution.
Hence, as the number of "primary" broken bonds at the 
peak load increases with increasing system sizes, 
the fracture strength distribution for larger lattice systems tends to be lognormal 
distribution.

\par
\vskip 0.70em%
\noindent
REMARK 1: In the case of narrowly distributed breaking thresholds, breaking 
of a bond significantly influences the subsequent bond breaking process, and the 
lattice system reaches its peak load soon after breaking fewer "primary" bonds. In 
particular, in the weakest-link hypothesis, the fracture strength of the system is 
identified with the breaking of the first bond, and hence the fracture strength 
distribution is based on extreme-value theory. Qualitatively, the assumption of broadly distributed 
breaking thresholds is supposed to distinguish the scenario of large number of 
"primary" broken bonds from the case of fewer "primary" broken bonds 
before the system reaches peak load. It is in this sense that we use the notion of broadly 
distributed heterogeneities.

\section{Numerical Simulations}
In this study, we pursue an electrical equivalence to the mechanical problem 
\cite {delaplace,deArcangelis89,orbach83}. 
We assume equivalence between
electrical current, voltage, and conductance in the
electrical system and the mechanical stress, strain, and Young's modulus
in the mechanical system, respectively. Similarly, we also assume an equivalence
in the breakdown process, that is, equivalence between successive burning of fuses
leading to the loss of electrical network conductivity and the mechanical breaking of
bonds leading to lattice system failure.
The main advantage of modeling a mechanical problem using an electrical analogy
is that the number of degrees of freedom in the system is significantly reduced, 
thereby increasing the system size that can be simulated using the same computational
power.

\par
Consider a two-dimensional lattice of size $L \times L$. The 
model adopted in this work is similar to the model III ({\it random threshold}) presented in 
Ref. \cite{sahimi86}, however, as mentioned in Section 2, we start 
with a fully intact lattice system and break one bond at a time until the 
lattice system falls apart. In this work, numerical simulations
are performed on triangular and diamond lattice topologies with periodic boundary
conditions in the horizontal direction.
The elastic response of each bond in the lattice is linear up to an
assigned threshold value at which brittle failure of the bond occurs. The disorder in the
system is introduced by assigning random maximum threshold current values, $i_c$,
(which is equivalent to the breaking stress in mechanical problem) to each of the fuses (bonds)
in the lattice,
based on an assumed probability distribution.
The electrical conductance (stiffness in the mechanical problem) is assumed to be
the same and equal to unity for all the bonds in the lattice. This is justified since the
conductance (or stiffness) of a heterogeneous solid converges rapidly to its scale
independent continuum value. The probability distribution of failure thresholds
is
dependent on the particular type of material considered. However, since our focus here is
not on modeling any specific type of material, but on the generic features of
damage evolution in disordered systems,
we choose, a uniform probability distribution, which is constant between
0 and 1. A broad thresholds distribution represents large disorder and
exhibits diffusive damage (uncorrelated burning of fuses) 
leading to progressive damage localization, whereas a very narrow
thresholds distribution exhibits brittle failure in which a single crack propagation
causes material failure. 

\par
Periodic boundary conditions are imposed in the horizontal direction to simulate
an infinite system and a constant voltage difference (displacement or strain)
is applied between the top and
the bottom of lattice system. The simulation is initiated with a lattice of intact
fuses in which disorder is introduced through random breaking thresholds. The voltage
$V$ across the lattice system is increased until a fuse (bond breaking) burns out.
The burning of a fuse occurs whenever the electrical current (stress)
in the fuse (bond) exceeds the
breaking threshold current (stress) value of the fuse. The current is redistributed
instantaneously after a fuse is burnt. The voltage is then gradually increased until
a second fuse is burnt, and the process is repeated.

\par
\begin{figure}[hbtp]
\centerline{\includegraphics[width=12cm]{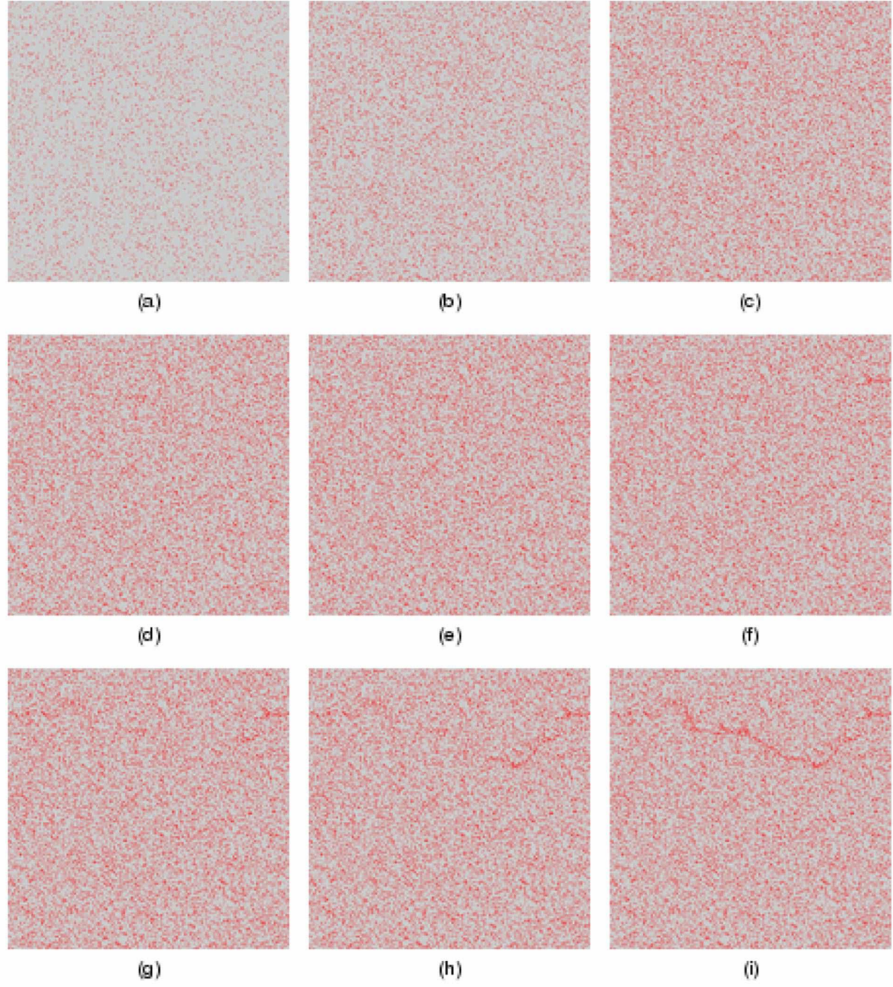}}
\caption{Snapshots of damage in a typical triangular lattice system of size $L = 512$. Number of broken bonds at the peak load and at failure are 83995 and 89100, respectively. (a)-(i) represent the snapshots of damage after breaking $n_b$ number of bonds. (a) $n_b = 25000$ (b) $n_b = 50000$ (c) $n_b = 75000$ (d) $n_b = 80000$ (e) $n_b = 83995$ (peak load) (f) $n_b = 86000$ (g) $n_b = 87000$ (h) $n_b = 88000$ (i) $n_b = 89100$ (failure)}
\label{fig1a}
\end{figure}

The above choice of redistributing the
current after breaking a fuse assumes that the current relaxation in the
lattice system is much faster than the breaking of a fuse.
Thus, each time a
fuse is burnt, it is necessary to calculate the current redistribution
in each of the fuses in the lattice. This is very time consuming, especially with
increasing lattice system size. The authors have developed a multiple-rank Cholesky updating 
algorithm for modeling relaxation processes in disordered systems \cite{nukala03}. 
In comparison with the most sophisticated Fourier accelerated iterative schemes 
used for modeling lattice breakdown \cite{batrouni88,batrouni98}, 
this algorithm significantly reduced the computational time required for solving large 
lattice systems. Figure \ref{fig1a} presents the snapshots of damage 
evolution for the case of a uniformly distributed random thresholds model problem
in a triangular lattice system of size $L = 512$. Based on these snapshots 
(Fig. \ref{fig1a}(a)-(e)), it is clear that the bond breaking occurs more or less 
randomly until very close to the peak load.
Since the response of the lattice system based on the
above numerical algorithm corresponds to a specific realization of random bond breaking
thresholds, an ensemble averaging of the numerical results is necessary to obtain
a realistic representation of the lattice system response.
Table 1 presents the number of configurations, $N_{config}$,
over which statistical averaging is
performed for different lattice sizes.

\section{Results and Discussion}

\subsection{Distribution of Fracture Strengths}
Conventionally, Weibull and Gumbel distributions are used to fit the fracture strength
data for brittle materials \cite{jayatilaka,orlovskaja}.
However, as Weibull mentioned in his pioneering paper, the Weibull distribution should be considered 
as an empirical one on an equal footing with other type of distributions \cite{weibull,orlovskaja}. 
In material 
science applications, lognormal, power law, Gamma, Type-I extreme value, and bimodal distributions 
are also often used for describing the fracture strength distribution 
\cite{orlovskaja,sigl,lissart,peterlik1,chao}.

\par
In randomly diluted disordered elastic networks, a Gumbel distribution better fits the fracture strengths
distribution far away from the percolation threshold and a Weibull distribution provides
a better fit close to the percolation threshold. 
These ideas are generally based on
the functional form of probability density of the defect clusters
that are obtained from broken bond
cluster statistics \cite{chakrabarti,sahimi98,duxbury86,duxbury87,duxbury88}.
In these randomly diluted disorder problems, the defect cluster size
distribution is exponential far
away from the percolation threshold and follows a power law close to the percolation
threshold. 
An exponential defect clusters distribution leads to Gumbel distribution
for fracture strengths and a power-law distribution of defect clusters leads to
Weibull form for fracture strengths.
\par
Duxbury et al \cite{duxbury86} 
studied the distribution of fracture thresholds in the disordered
media with randomly diluted bonds. Subsequently, this study was applied to a variety of 
lattice models with different local behavior including central-force spring models
\cite{beale88,sahimi933}, 
Born models \cite{hassold89}, and bond-bending models \cite{sahimi933}. 
In all these studies, fracture strength of the 
lattice system was defined as the stress required for breaking the very first bond 
in the system. 
These studies along with the analytical investigations based on largest crack size 
\cite{kahng}
concluded that the fracture threshold decreased with the system 
size $L$ as a power law of $log(L)$ and the fracture threshold distribution 
is best described by a double exponential (modified Gumbel) distribution. 
However, numerical 
simulations on intact two-dimensional triangular and diamond lattice topologies 
for heterogeneous materials with a broad disorder
indicate 
that the entire lattice network does not fall apart as soon as the first bond is 
broken. In fact, the number of broken bonds at the peak load scales as a power law. 
This suggests that the arguments used for deriving the breaking strengths of 
the first bond may not be extended to this type of problems.

\par
In this study, fracture strength of a lattice system is defined as the 
stress corresponding to the 
peak load of the lattice system
response. 
A schematic of a typical lattice system response for a given distribution of random bond breaking 
thresholds is shown in Fig. \ref{fig1}. Each realization of the random bond 
breaking thresholds 
results in a specific fracture strength value. The
distribution of these 
fracture strengths sampled over an ensemble of configurations is the subject of interest 
in this section. In the following, we investigate the validity of Weibull and Gumbel distributions
to represent the fracture strength data corresponding to the peak load of the lattice 
response.
 
\par
\begin{figure}[hbtp]
\begin{tabular}{cc}
{\includegraphics[width=12cm]{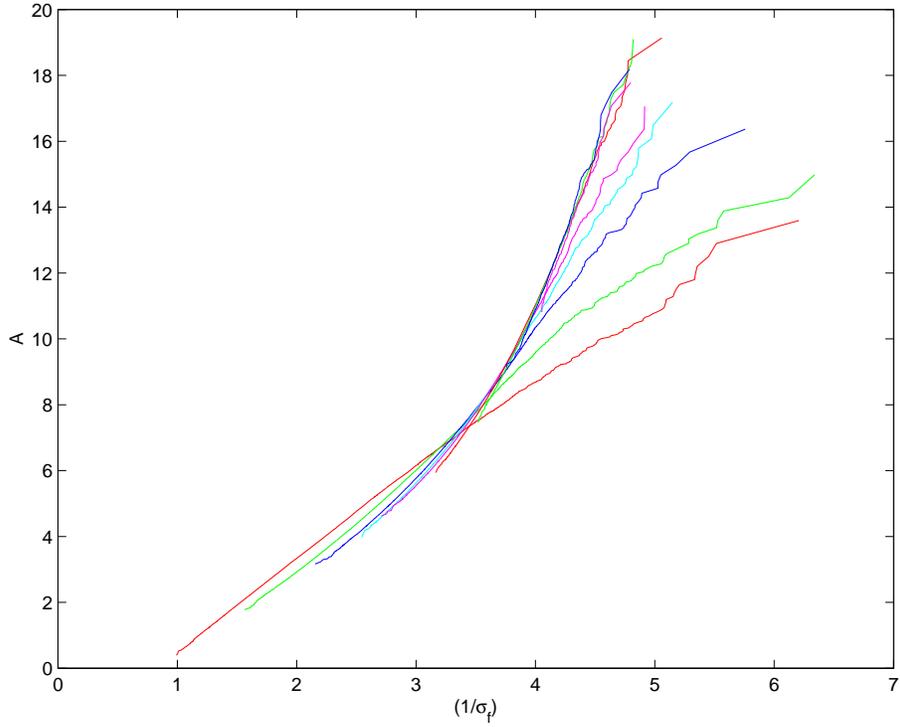}} \\
(a) \\
{\includegraphics[width=12cm]{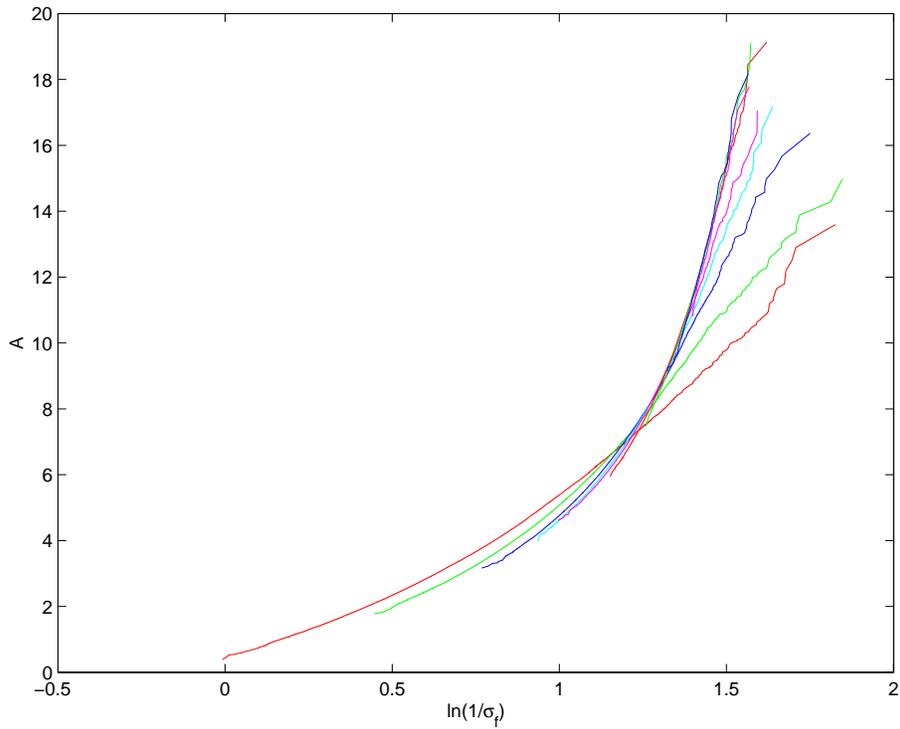}} \\
(b)
\end{tabular}
\caption{Probability distribution fits for fracture strengths at the peak load response 
in a triangular lattice for different lattice system sizes $L$ = \{4, 8, 16, 24, 32, 64, 128, 256, 512\}. 
(a) Gumbel distribution (b) Weibull distribution}
\label{fig3}
\end{figure}

The Gumbel distribution for fracture strengths $\sigma_f$ is given by
\begin{equation}
P_G(\sigma_f) = 1 - exp(-c L^d ~exp(-\frac{k}{\sigma_f^\delta})) \label{gumbel}
\end{equation}
and the Weibull distribution is 
\begin{equation}
P_W(\sigma_f) = 1 - exp(-c L^d \sigma_f^m) \label{weibull}
\end{equation}
where $k$, $\delta$, $c$ and $m$ are constants, and $d$ denotes the lattice dimension.
The validity of Gumbel and 
Weibull distributions to the fracture strength data can be tested by rewriting 
Eqs. (\ref{gumbel}) and (\ref{weibull}) as
\begin{eqnarray}
A & = & k ~ \left(\frac{1}{\sigma_f^\delta}\right) ~-~ ln ~c \label{gda}
\end{eqnarray}
for the Gumbel distribution and
\begin{eqnarray}
A & = & m ~ ln\left(\frac{1}{\sigma_f}\right) ~-~ ln ~c \label{wda}
\end{eqnarray}
for the Weibull distribution. In Eqs. (\ref{gda}) and (\ref{wda}), the variable $A$ is 
defined as
\begin{eqnarray}
A & = & -ln\left[-\frac{ln\left(1 - P(\sigma_f)\right)}{L^d}\right] \label{eqnA}
\end{eqnarray}
where $P(\sigma_f)$ is $P_G(\sigma_f)$ in the case of Gumbel distribution, and 
$P_W(\sigma_f)$ for Weibull distribution.
Figure \ref{fig3}(a) presents Gumbel fit for the fracture strength 
distribution for triangular lattice network using Eq. (\ref{gda}). Similarly, 
Fig. \ref{fig3}(b) presents the Weibull distribution fit (Eq. (\ref{wda})) 
for the fracture strengths 
obtained using simulations on triangular lattices. From these figures, it is 
clear that fracture strength data obtained for two different lattice sizes 
does not align onto a single straight line as it should, if the data were to follow
Eq. (\ref{gda}) or (\ref{wda}). 
The fracture strength results based on diamond lattice network exhibit similar 
trends indicating that in the case of highly disordered materials,
neither Gumbel nor
Weibull distributions may represent the fracture strengths distribution accurately.

\par
\begin{figure}[hbtp]
\begin{tabular}{cc}
{\includegraphics[width=8cm]{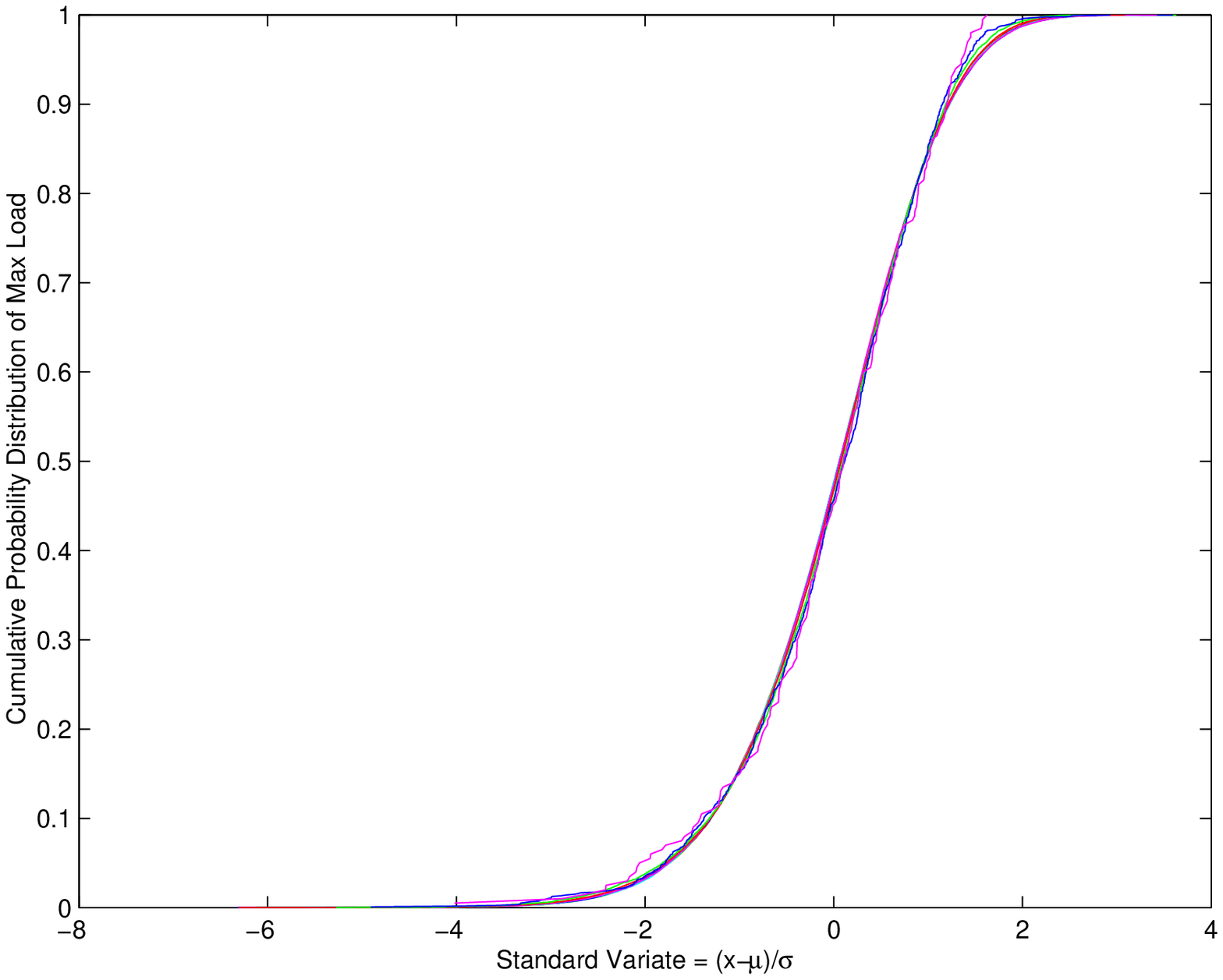}} &
{\includegraphics[width=8cm]{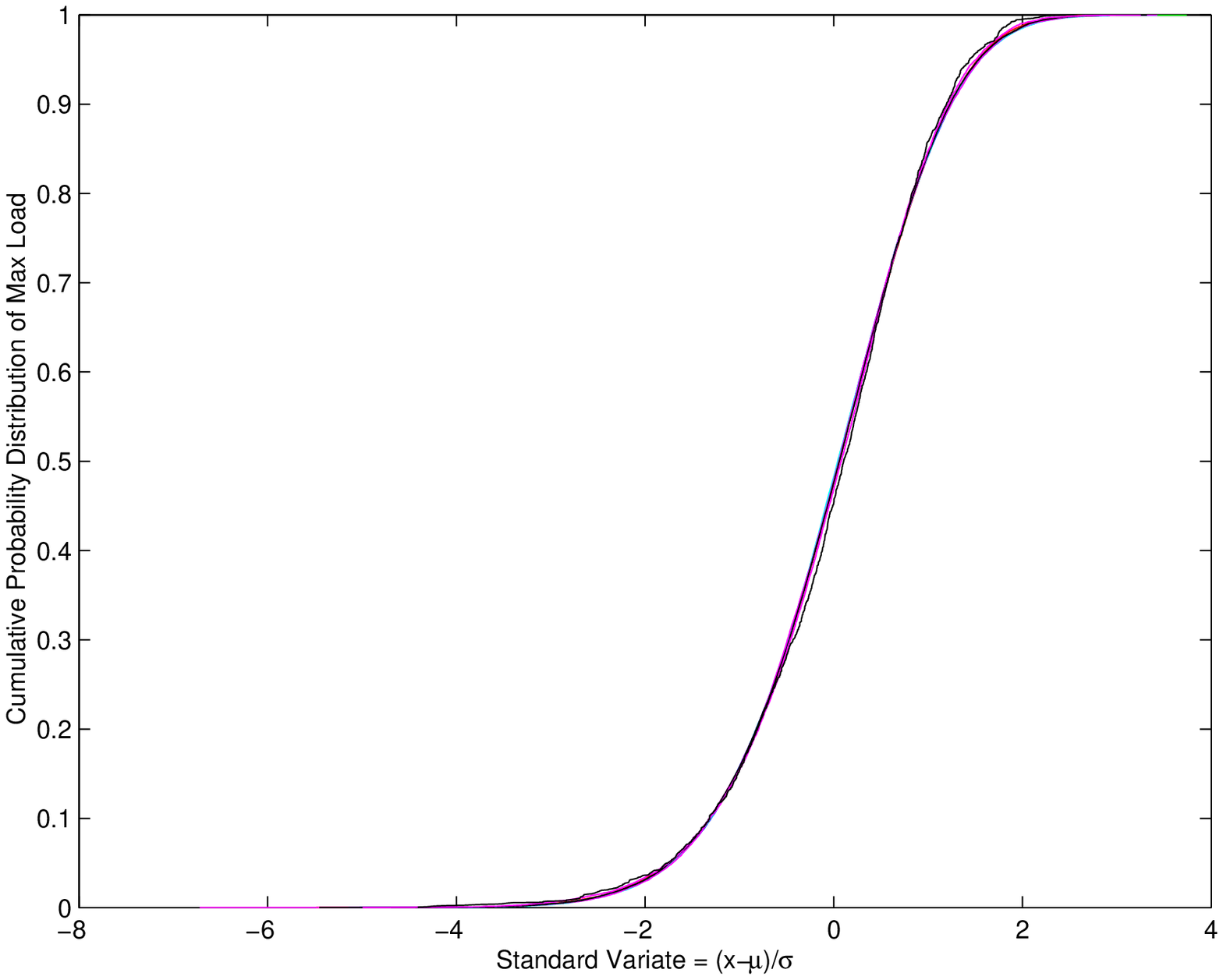}} \\
(a) & (b) \\
{\includegraphics[width=8cm]{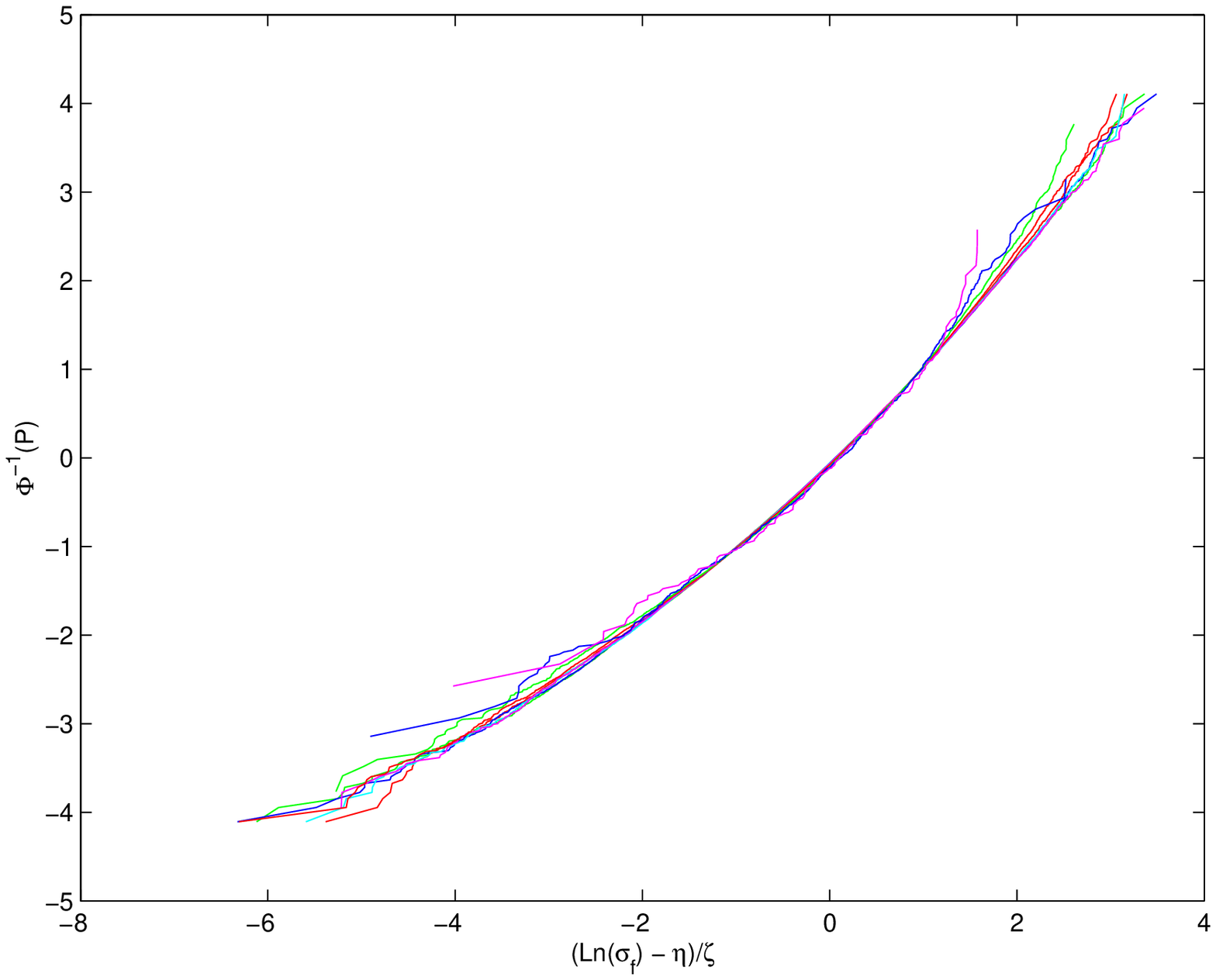}} &
{\includegraphics[width=8cm]{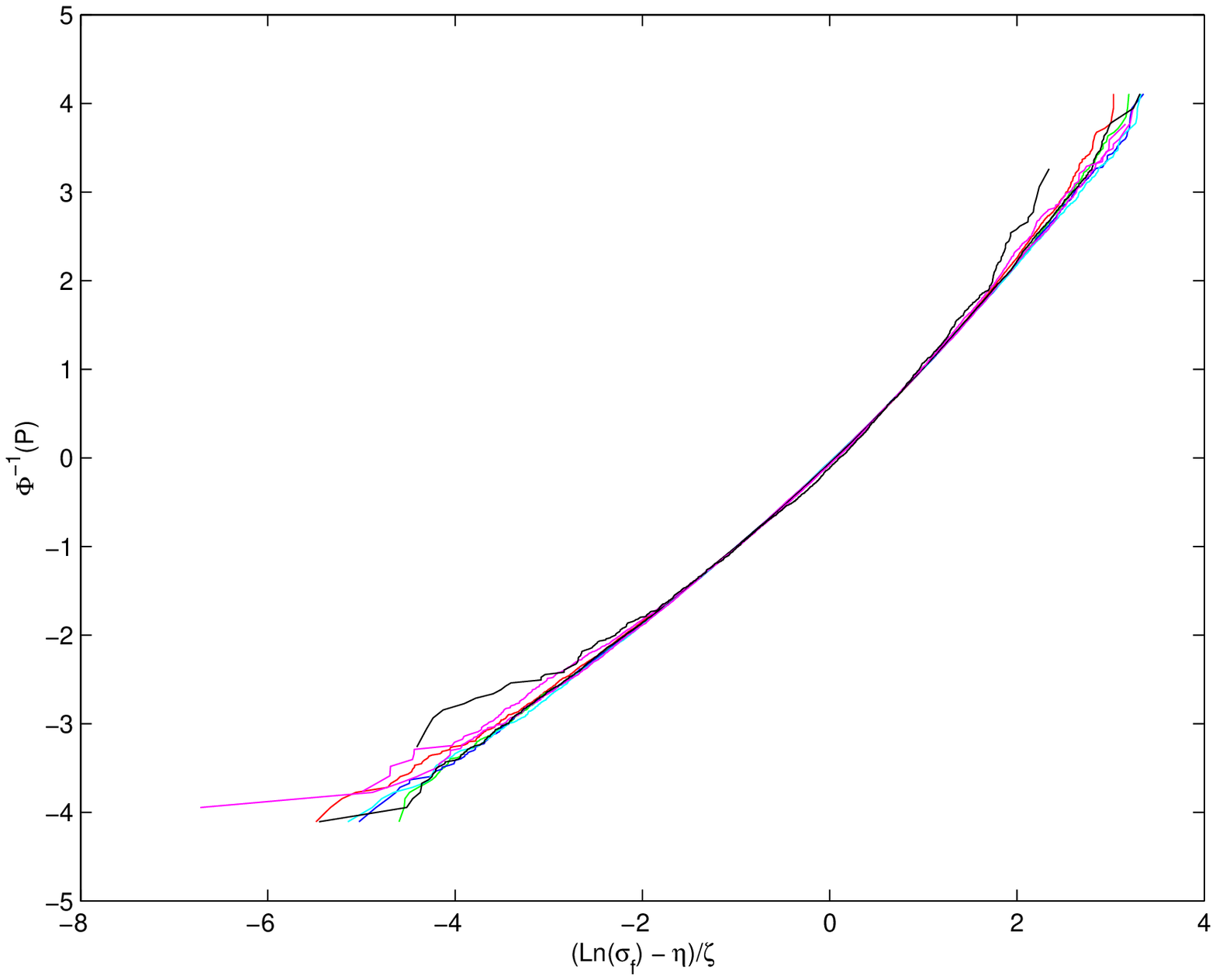}} \\
(c) & (d) 
\end{tabular}
\caption{Lognormal distribution fit for fracture strengths at the peak load response 
(a) triangular lattice (b) diamond lattice. 
Reparametrized lognormal distributions: (c) triangular lattice (d) diamond lattice. 
For triangular systems, $L$ = \{4, 8, 16, 24, 32, 64, 128, 256, 512\}, and 
for diamond lattices, $L$ = \{4, 8, 16, 24, 32, 64, 128, 256\}}
\label{fig5}
\end{figure}

The first main result of this study is that 
for materials with broadly distributed heterogeneities,
a lognormal distribution
represents the fracture strength of the macroscopic system more adequately than
previously used (modified) Gumbel and Weibull distributions.
Lognormal distribution has been used in engineering practice to represent the fracture 
strengths of materials \cite{wolstenholme,own}.
The lognormal distribution can be understood to have evolved as a 
consequence of multiplicative nature of large number of 
random distributions representing the individual scale factors necessary to 
break the subsequent "primary" bonds leading up to the peak load. 
The precise character of the individual distributions that are 
multiplied to give the final distribution is irrelevant as long as the number of 
"primary" broken bonds up to the peak load is large. 
Figures \ref{fig5}(a) and \ref{fig5}(b) present the
cumulative fracture strength versus the standard lognormal variable, $\xi$, defined as
$\xi = \frac{Ln(\sigma_f) - \eta}{\zeta}$,
for triangular and diamond lattice networks respectively.
In the above description, $\eta$ and $\zeta$
refer to the mean and the standard deviation of the logarithm of $\sigma_f$.
These figures indicate that the fracture strength distribution collapses onto a 
single curve for different lattice system sizes, which is an improvement compared to 
(modified) Gumbel and Weibull distributions.
A better representation to test the lognormal description for fracture strengths 
is to plot the inverse of the cumulative probability, $\Phi^{-1}(P(\sigma_f))$,
against the standard lognormal variable, $\xi$. In the above description, 
$\Phi( \cdot )$ denotes the standard normal probability function.
Figures \ref{fig5}(c) and \ref{fig5}(d) present the lognormal fit for the
cumulative fracture strength
distributions obtained for triangular and diamond lattice networks respectively.
From these figures, it is clear that the fracture strength distribution
obtained for different lattice system sizes collapses onto a single curve, 
albeit minute deviation from straight line behavior is evident. We have also 
used the normal distribution to collapse the fracture strength data of triangular and 
diamond lattice systems. Although the data collapse is reasonable, it is not as good as 
that of lognormal distribution.

\subsection{Mean Fracture Strength}
\begin{figure}[hbtp]
\begin{tabular}{cc}
{\includegraphics[width=12cm]{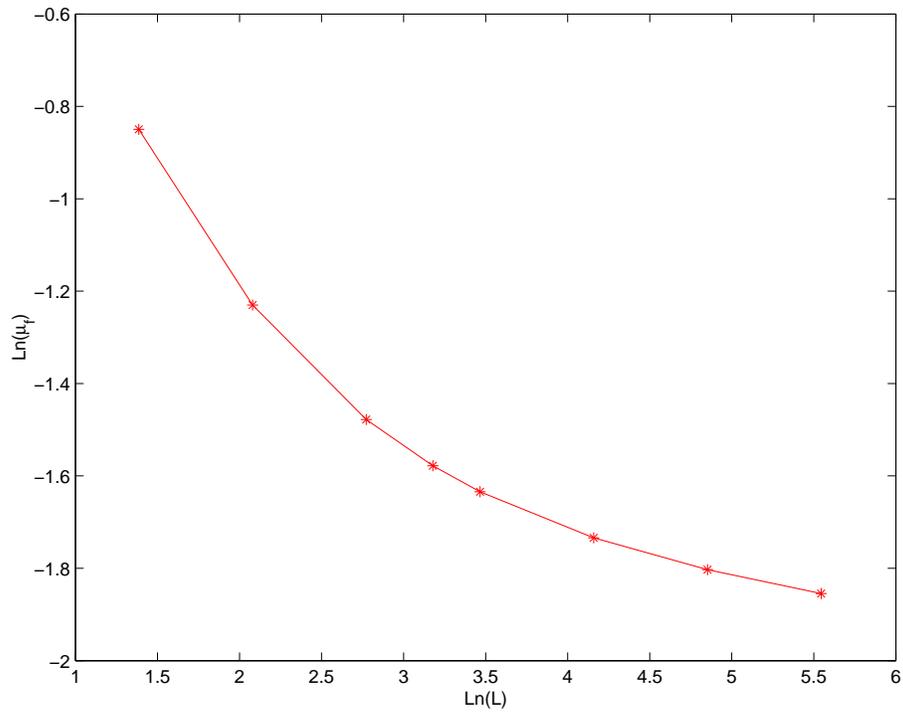}} \\
(a) \\
{\includegraphics[width=12cm]{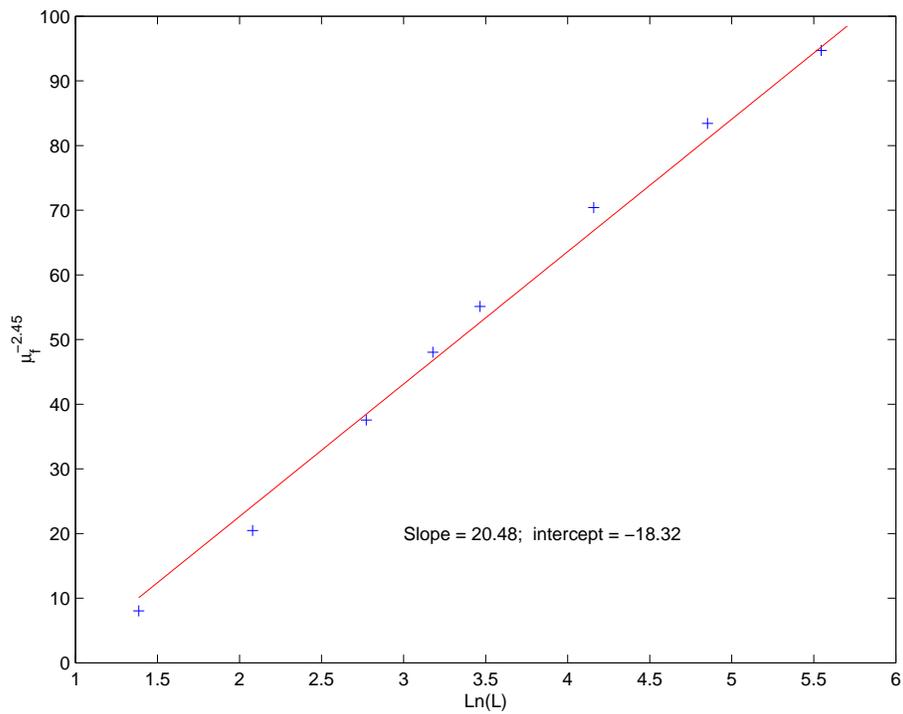}} \\
(b)
\end{tabular}
\caption{Mean fracture strength versus diamond lattice system size. 
(a) Weibull fit based on Equation (\ref{wsigsc}) (b) Modified Gumbel fit 
based on Equation (\ref{gsigsc})}
\label{fig9}
\end{figure}

The second main result of this study is concerned with the scaling law for 
the mean fracture strength of the lattice system. Table 1 presents the 
mean fracture strength data for various triangular and diamond lattice system sizes.
Cumulative fracture strength distributions may be used to derive the scaling form
of mean fracture strength. For example, when the Weibull distribution represents the 
fracture strengths, the mean fracture strength, $\mu_f$, scales as a power law, i.e.,
\begin{eqnarray}
\mu_f & \sim & L^{-\frac{d}{m}} \label{wsigsc}
\end{eqnarray}
Similarly, in the case of fracture strengths represented by Gumbel distribution, 
we have
\begin{eqnarray}
\mu_f^\delta & = & \frac{k}{(ln ~c - a_1) ~+~ d ~ln ~L} \nonumber \\
& = & \frac{1}{A_1 ~+~ B_1 ~ln ~L} = \frac{1}{ln (A_2 ~L^{B_1})} \label{gsigsc}
\end{eqnarray} 
where $a_1$, $A_1$, $B_1$, and $A_2 = exp(A_1)$ are constants that are related to the 
parameters $k$ and $c$ of the Gumbel distribution. Equation (\ref{gsigsc}) is 
same as the relation proposed by \cite{duxbury88,beale88} that shows the effect of 
sample size $L$ on average breaking voltage or failure strength. 
Figures \ref{fig9}(a) and \ref{fig9}(b) present the mean fracture strength $\mu_f$ versus
the lattice size $L$ based on the form of Eqs. (\ref{wsigsc}) and (\ref{gsigsc}), 
respectively, for diamond lattice system. The results for triangular lattice systems 
exhibit similar behavior. From these figures, it is evident that mean fracture strength does not
follow a power-law dependence on the lattice size $L$. The results presented in 
Fig. \ref{fig9}(b) indicate that Eq. (\ref{gsigsc}) may be able to represent the 
mean fracture strengths reasonably well even though the Gumbel distribution is 
inadequate to represent the cumulative fracture strength 
distribution. The value of the exponent $\delta$ is approximately equal to $2.45$ and is 
consistent with the values reported in the literature \cite{duxbury88,beale88}.
The coefficient $B_1$ (slope in Fig. \ref{fig9}(b)) is large ($\approx 20.5$) 
and is, once again, in agreement with the divergence behavior expected as the 
lattice system approaches failure \cite{duxbury88,beale88}.
This weak (logarithmic) dependence of mean fracture strength on system
size $L$ is consistent with the analytical and previously reported numerical 
results on randomly diluted 
networks \cite{chakrabarti,sahimi98,duxbury86,duxbury87,duxbury88,beale88}.

\par
\begin{figure}[hbtp]
\centerline{\includegraphics[width=12cm]{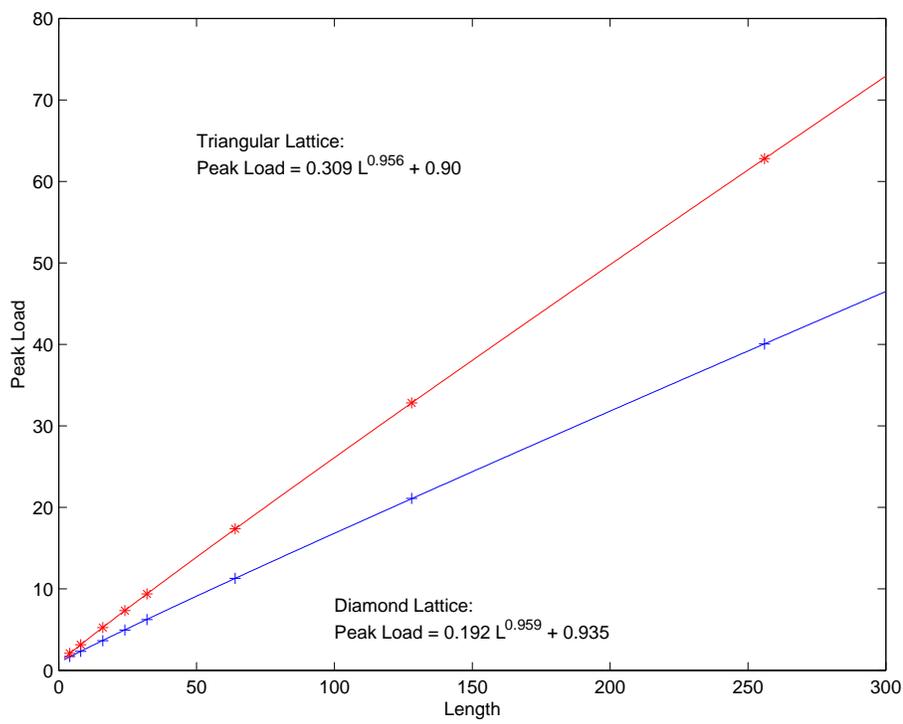}}
\caption{Average peak load versus the lattice system size, $L$}
\label{fig11}
\end{figure}

Alternatively, we have plotted the ensemble averaged peak load, ${\bar{F}}_{peak}$,
versus the lattice system size, $L$, as shown in Fig. \ref{fig11}. 
From the 
numerical simulation results presented in Fig. \ref{fig11}, it is clear that the
ensemble averaged peak load ${\bar{F}}_{peak}$ for triangle and diamond lattice
topologies may be expressed as
\begin{eqnarray}
{\bar{F}}_{peak} & = & C_0 ~L^{\alpha} ~+~ C_1 \label{Fpeak}
\end{eqnarray}
where $C_0$ and $C_1$ are constants. Thus, the mean fracture strength, $\mu_f$, 
defined as $\mu_f = \frac{{\bar{F}}_{peak}}{L}$, is given by
\begin{eqnarray}
\mu_f & = & C_0~L^{\alpha - 1} ~+~ \frac{C_1}{L} \label{muf}
\end{eqnarray}
The results shown in Fig. \ref{fig11} indicate that for both the 
triangular and diamond lattice topologies, the exponent $\alpha$ in 
Eq. (\ref{Fpeak}) is approximately equal to {\it 0.96}. This, in 
turn results in a very small exponent value equal to {\it -0.04} in the 
first term of the Eq. (\ref{muf}). A very small negative value of the 
exponent $(\alpha - 1)$ is equivalent to a logarithmic correction, i.e., 
for $(1-\alpha) << 1$, $L^{\alpha - 1} \sim (log(L))^{-\psi}$. Thus, an 
alternative expression for the mean fracture strength may be expressed as
\begin{eqnarray}
\mu_f & = & \frac{\mu_f^\star}{(Log L)^\psi} ~+~ \frac{c}{L} \label{mufRGmod}
\end{eqnarray}
where $\mu_f^\star$ and $c$ are constants that are related to the constants 
$C_0$ and $C_1$ of Eq. (\ref{muf}). 
This shows that the mean fracture strength of the lattice system decreases very slowly with 
increasing lattice system size, and scales as $\mu_f \approx \frac{1}{(Log L)^\psi}$ for 
very large lattice systems. 

\section{Conclusions}
This paper presents a theoretical investigation supplemented by 
numerical simulations to describe the fracture strength distribution 
of a lattice system. The discrete lattice system considered is fully intact 
at the beginning of the analysis and the damage is accumulated progressively
by breaking one bond at a time until the entire lattice system falls apart.
The fracture strength of a lattice system is defined as the stress corresponding
to the peak load of the lattice system response. This is in contrast with the
earlier works, wherein either randomly diluted lattice systems close to the
percolation threshold were considered or the stress required to break the very first
bond was defined as the fracture strength of the lattice system. 
\par
Our study presents two main results. 
First, for materials with broadly distributed heterogeneities, 
a lognormal distribution 
represents the fracture strength of the macroscopic system more adequately than 
previously used distributions such as (modified) Gumbel and Weibull. 
The lognormal distribution can be understood to have evolved as a 
consequence of multiplicative nature of large number of 
random distributions representing the stress scale factors necessary to break 
the subsequent "primary" bonds leading up to the peak load.
The precise character of these individual distributions that are 
multiplied to give the fracture strength distribution is irrelevant as long as the number of 
"primary" broken bonds up to the peak load is large. 
Hence, as a consequence of the central limit theorem, 
the system fracture strength probability distribution approaches a lognormal distribution.
Numerical simulations based on two-dimensional triangular and diamond lattice 
topologies substantiate that a lognormal distribution represents an excellent fit 
for the fracture strength distribution. 
\par
Second, the mean fracture strength of the 
lattice system behaves as $\mu_f = \frac{\mu_f^\star}{(Log L)^\psi} ~+~ \frac{c}{L}$, 
and scales as $\mu_f \approx \frac{1}{(Log L)^\psi}$ 
as the lattice system size, $L$, approaches infinity. 

\par
\vskip 1.00em%
\noindent
{\bf Acknowledgment} \\
This research is sponsored by the Mathematical, Information and Computational Sciences
Division, Office of Advanced Scientific Computing Research, U.S. Department of Energy under
contract number DE-AC05-00OR22725 with UT-Battelle, LLC. The first author wishes to thank 
Dr. Muhammad Sahimi for many helpful suggestions on the manuscript.

\newpage

\bibliography{damage}
\bibliographystyle{unsrt}

\newpage

\begin{table}[hbtp]
  \leavevmode
  \begin{center}
  \caption{Peak Load}
  \vspace*{1ex}
  \begin{tabular}{|c|c|c|c|c|c|}\hline
  L  & $N_{config}$ & \multicolumn{2}{c|}{Triangular} & \multicolumn{2}{c|}{Diamond}\\\cline{3-6}
     & & Mean & Std & Mean & Std\\
  \hline
  4 & 50000 & 2.087 & 0.4328 & 1.710 & 0.4248 \\
  8 & 50000 & 3.148 & 0.4545 & 2.338 & 0.4209 \\
 16 & 50000 & 5.261 & 0.518 & 3.650 & 0.4485 \\
 24 & 50000 & 7.336 & 0.5822 & 4.953 & 0.4865 \\
 32 & 50000 & 9.377 & 0.6434 & 6.244 & 0.5219 \\
 64 & 50000 & 17.372 & 0.8595 & 11.301 & 0.6611 \\
128 & 12000 & 32.82 & 1.2952 & 21.09 & 0.9045 \\
256 & 1200 & 62.79 & 2.0251 & 40.06 & 1.380 \\
512 & 200 & 120.49 ($121.13^\star$) & 3.5789 & & \\
  \hline
  \end{tabular}
\\ {$\star$ estimated value based on the equation given in the inset of Fig. 11}
  \label{table1}
  \end{center}
\end{table}
\newpage

\newpage

\newpage

\newpage

\newpage

\newpage

\end{document}